\documentclass[epj]{svjour}
\usepackage{graphics}
\usepackage{dcolumn}
\usepackage{amsfonts}
\usepackage{amsmath}
\usepackage{amssymb}
\usepackage{bm}

\newif\ifpdf
\ifx\pdfoutput\undefined
\pdffalse % we are not running PDFLaTeX
\else
\pdfoutput=1 % we are running PDFLaTeX
\pdftrue \fi \ifpdf
\usepackage[pdftex]{graphicx}
\else
\usepackage{graphicx}
\fi

\newcommand{\uv}{{\mathbf u}}
\newcommand{\nv}{{\mathbf n}}

\newcommand{\Ha}{{\mathcal H}}

\newcommand{\brm}[1]{\bm{{\rm #1}}}

\begin{document}

\title{Commentary on ``Mechanical properties of mono-domain side
chain nematic elastomers" by P. Martinoty {\em et al}.}

\author{Olaf Stenull and T. C. Lubensky
}

\institute{Department of Physics and Astronomy, University of
Pennsylvania, Philadelphia, Pennsylvania 19104, USA}

\date{Received: date / Revised version: date}

\abstract{
We discuss the rheology experiments on nematic elastomers
by Martinoty {\em et al}.\ in the light of theoretical
models for the long-wavelength low-frequency dynamics of
these materials. We review these theories and discuss how
they can be modified to provide a phenomenological
description of the non-hydrodynamic frequency regime probed
in the experiments. Moreover, we review the concepts of
soft and semi-soft elasticity and comment on their
implications for the experiments.
}

\authorrunning{O. Stenull and T. C. Lubensky
}

\titlerunning{Commentary on nematic elastomers}

\maketitle

Nematic elastomers are crosslinked rubbers with the
uniaxial symmetry of a nematic liquid crystal
\cite{WarnerTer2003}.  One could argue, therefore, that
they are simply uniaxial rubbers.  A nematic elastomer can,
however, be distinguished from a simple uniaxial elastic
medium by the fact that, at least in an idealized limit, it
can form via a spontaneous phase transition from an
isotropic phase. Since this phase transition breaks the
continuous rotational symmetry of the isotropic phase, it
has an associated Goldstone mode whose manifestation is the
vanishing of the elastic modulus $C_5$ measuring the
elastic energy of strains in planes containing the
anisotropy axis \cite{GolLub89,Olmsted94}. Thus, the ideal
nematic elastomer has a ``soft" elasticity compared to a
traditional uniaxial elastic medium in which $C_5$ is
nonzero. In practice, ideal monodomain nematic elastomers
do not form in the absence of some aligning field, which is
usually produced by weakly crosslinking a sample in the
isotropic phase, stretching to produce a uniaxial
configuration, and then crosslinking again \cite{FinKun97}.
The resultant material is weakly anisotropic in the
high-temperature paranematic phase and more anisotropic in
the low-temperature nematic phase. Since the
low-temperature phase arises from an already aligned phase
and does not break rotational symmetry, it does not exhibit
ideal soft elasticity. Rather, it exhibits ``semi-soft"
elasticity in which $C_5$ is small but nonzero. Thus, for
small strains, a semi-soft nematic elastomer is truly a
uniaxial solid. At higher strains, however, it exhibits
properties, such as a nearly constant stress for increasing
strain, characteristic of a soft nematic elastomer \cite{warner_99}.

The static elastic properties and phase transitions of both
soft and semi-soft nematic elastomers are reasonably well
understood~\cite{WarnerTer2003,GolLub89,LubenskyXin2002,stenull_lubensky_epl2003,stenull_lubensky_anomalousNE_2004},
provided the effects of random stresses~\cite{GolLub89,Xing_Radz_03}
can be ignored. They can be described either in terms of models involving
strain only or in terms of models with coupling between a
traditional Maier-Saupe-de-Gennes nematic
symmetric-traceless order parameter $Q_{ij}$ and strain.
Considerable progress has also been made toward
understanding the dynamics of nematic elastomers
\cite{brand_pleiner_1994,terentjev&Co_NEhydrodyn,stenull_lubensky_2004}
again in the limit in which random stresses and fields can
be ignored. As in a standard elastic medium, purely
hydrodynamical equations, which describe all modes with
frequencies $\omega$ smaller than the smallest
characteristic inverse decay time $\tau^{-1}$ when
wavenumber $q$ tends to zero, involve only the displacement
field $\uv$ and not the nematic director $\nv$, which
relaxes to the local stain in a nonhydrodynamic time
$\tau_n$.  Even though the nematic director is not strictly
speaking a hydrodynamic variable in either soft or
semi-soft nematic elastomers, it is of some interest to
develop phenomenological equations describing both nematic
director and elastic displacement. Two distinct
derivations~\cite{WarnerTer2003,terentjev&Co_NEhydrodyn,stenull_lubensky_2004}
lead to identical sets of coupled director-displacement
equations, which we will refer to as the NED
(nematic-elastomer dynamic) equations, that reduce to the
rigorous hydrodynamic equations in terms of displacement
only at frequencies $\omega \tau_n \ll 1$ and to the
standard equations of nematohydrodynamic  when elastic
rigidity is turned off.

Rheological measurements of the complex modulus~\cite{footnote1}
\begin{align}
G^\ast (\omega )= G^\prime (\omega) - i G^{\prime \prime}( \omega)
\end{align}
provide a useful experimental probe of the dynamical
properties of viscoelastic systems. They typically probe a
wide range of frequencies that stretch from the
low-frequency hydrodynamic regime to well above it.  Thus,
interpreting these experiments requires theoretical models
of dynamics beyond hydrodynamics. A nematic elastomer is
characterized in general by relaxation times associated
with director relaxation and with other modes, which we
will simply refer to as elastomer modes. In their simplest
form, the NED equations assume the director relaxation time
$\tau_n$ is the longest non-hydrodynamic relaxation time,
and that it is well separated from the longest elastomer
time $\tau_E$. They describe dynamics for both $\omega
\tau_n \ll 1$ and $\omega \tau_n >1$ and provide
non-trivial predictions about $G^* ( \omega)$, in
particular the appearance of a plateau in $G'( \omega )$
and an associated dip in $G''(\omega)$ for $\tau_n^{-1} <
\omega < \tau_E^{-1}$.
Experiments~\cite{clarke&Co_2001,stein&Co_2001,hotta_terentjev_2003,terentjev&Co_2003,martinoty&Co_2004}
to date do not exhibit the plateau predicted by the NED
equations. This has led Martinoty {\em et al}.\ to question
their validity.

Here we will review both the hydrodynamical and NED
equations for nematic elastomers and discuss how they can
be modified to provide a phenomenological description of
the non-hydrodynamic frequency regime probed in rheology
experiments. Moreover, we review the concepts of soft and
semi-soft elasticity and comment on their implications for
rheology experiments.

We begin with a review of the dynamics of traditional
isotropic solids, of which rubber is a particular example.
The elastic free energy of an isotropic solid is
characterized by a bulk modulus $B$ and a shear modulus
$\mu$, and dissipation is characterized by a bulk viscosity
$\zeta$ and a shear viscosity $\eta$, which are both
frequency independent at low frequencies. The complex shear
 modulus in the hydrodynamic limit is simply
\begin{equation}
G^*( \omega ) = \mu - i \omega \eta .
\label{eq:G*1}
\end{equation}
This form is rigorous in the hydrodynamic limit where
$\omega \tau \ll 1$. $\mu$ and $\eta$ can vary widely in
magnitude from system to system. Nothing more can be said
about the behavior of $G^*(\omega)$ without some more
microscopic model of the dynamics of a particular system.

The Rouse model~\cite{rubinstein_colby_2003} provides a
simple description of the dynamics of polymeric melts. The
dynamics of even isotropic elastomers and rubbers is far
more complex than that of melts, and the Rouse model does
not necessarily provide a good physical description of
their dynamical modes.  Nevertheless, experimental curves
for $G^*(\omega)$ in nematic elastomers are very similar to
those predicted by the Rouse model, showing in particular a
region in which $G^{\prime}( \omega )$ and $G^{\prime
\prime} ( \omega )$ track each other with a
$\omega^{\alpha}$ behavior with $\alpha \approx 1/2$. It is
thus legitimate to use Rouse-model predictions as fitting
curves for nematic elastomers, whether or not it provides a
microscopically correct description of their dynamics. This
is the point of view we will take here. We parameterize the
complex storage modulus in the isotropic phase of nematic
elastomers as
\begin{equation}
G^*(\omega) = G_0 [ 1 + a \, h_R (- i \omega \tau_E )],
\label{eq:G*2}
\end{equation}
where $G_0 = n_x k T$, with $n_x$ being the number density
of polymer strands between crosslinks, and $a =
\sum_{p=1}^N p^{-2}$. $h_R ( x ) = x\, f_R ( x )$ is a
Rouse function with
\begin{equation}
\label{Rousefunction}
f_R ( x ) = \frac{\sum_{p=1}^N (p^2 + x)^{-1}}{\sum_{p=1}^N
p^{-2}} \, .
\end{equation}
In the limit $N \to \infty$, the summations in
Eq.~(\ref{Rousefunction}) can be evaluated via the Poisson
summation formula with the result
\begin{align}
\label{RousefunctionAna}
h_R ( x ) = \frac{3}{\pi^2 \, \sinh (\pi \sqrt{x})} \left\{
\pi \sqrt{x}  \cosh (\pi \sqrt{x}) -  \sinh (\pi \sqrt{x})
\right\}  .
\end{align}
When $\omega \tau_E \ll 1$, this expression reduces to
Eq.~(\ref{eq:G*1}) with $\mu = n_x k T$, and $\eta = (\pi^2
/6)\mu \tau_E$. When $\tau_E^{-1} < \omega < \tau_m^{-1}$,
$G' ( \omega ) \sim G''(\omega) \sim \sqrt{\omega}$.

A uniaxial solid is characterized by an elastic energy
\begin{align}
\label{uniax1} \mathcal{H}_{\brm{u}}^R& = \int d^3 x \,
\bigg\{ \frac{C_1}{2} \, u^2_{zz} + C_2 \, u_{zz} u_{ii} +
\frac{C_3}{2} \, u_{ii}^2
\nonumber \\
& + C_4 \, u_{ab}^2 + C_5^R \, u_{az}^2 \bigg\} \, .
\end{align}
with five elastic constants, where $u_{ij}$ with $i,j =
x,y,z$ is the strain tensor and $a,b = x,y$. The
superscript $R$ on the elastic modulus $C_5^R$ for shears
in plane containing the $z$-axis indicates that it has been
renormalized by director fluctuations. In soft nematic
elastomers, $C_5^R$ is zero, and it is necessary to add
bending terms proportional to $(\partial^2 u_a )^2$ to
ensure stability in all directions.  In practice all single
domain nematic elastomers are semi-soft with $C_5^R > 0$,
and the bending terms can be neglected. Associated with
each of the five elastic constants is a viscosity $\eta_k$.
The complex storage moduli for shears in the planes
containing the $z$-axis and in the plane perpendicular to
the $z$-axis are, respectively,
\begin{align}
G_{\parallel}^* ( \omega ) & =  C_5^R - i \omega \eta_5^R\, ,
\\
G_{\perp}^* ( \omega ) & =  C_4 - i \omega \eta_4 .
\end{align}
As in isotropic rubbers, we can say nothing more about the
behavior of $G_{||}^* ( \omega ) $ and $G_{\perp}^*(
\omega)$ without some microscopic model of the dynamics.
Since there are surely elastomer modes in the anisotropic
phase of nematic elastomers, one could argue that they are
dominant and that $G_{||}^*$ and $G_{\perp}^*$ have exactly
the same form as $G$ in an isotropic system, Eq.~(\ref{eq:G*2}), with possibly different values of $G_0$, $a$ and $\tau_E$. Indeed, the data presented in Martinoty,
{\em et al}.\ appear to fit this model with $\tau_E \sim
10^{-2}$, considerably longer than the Rouse time in
traditional isotropic rubbers. Certainly, however, there
are extra director modes in nematic elastomers that are not
present in isotropic ones, and a complete model of dynamics
beyond hydrodynamics must include them.

A phenomenological approach to the dynamics of nematic
elastomers including the nematic director begins with an
elastic free energy that includes both strain and the
director.  Following de Gennes~\cite{de_Gennes-1}, the
energy $\mathcal{H}$ can be written as a sum of the Frank
free energy $\Ha_\nv$ for the director, the strain and
strain-director coupling energy
\begin{align}
\label{uniax}
\mathcal{H}_{\brm{u}, \brm{n}}& = \int d^3 x \, \bigg\{
\frac{C_1}{2} \, u^2_{zz} + C_2 \, u_{zz} u_{ii} +
\frac{C_3}{2} \, u_{ii}^2
\nonumber \\
& + C_4 \, u_{ab}^2 + C_5 \, u_{az}^2 + \frac{D_1}{2} \,
Q_a^2  + D_2  \, u_{za} Q_a \bigg\} ,
\end{align}
where
\begin{eqnarray}
\label{defQ} Q_a = \delta n_a - \frac{1}{2} \, (\partial_z
u_a - \partial_a u_z)  \, .
\end{eqnarray}
When $Q_a$ relaxes to its equilibrium value $ -(D_2/D_1)
u_{az}$ in the presence of strain $u_{az}$, the energy
$\mathcal{H}_{\brm{u}, \brm{n}}$ reduces to $\Ha^R_{\uv}$
with $C_5^R = C_5 - D_2^2/(2 D_1)$.

\begin{figure}
\begin{center}
\includegraphics[width=8.0cm]{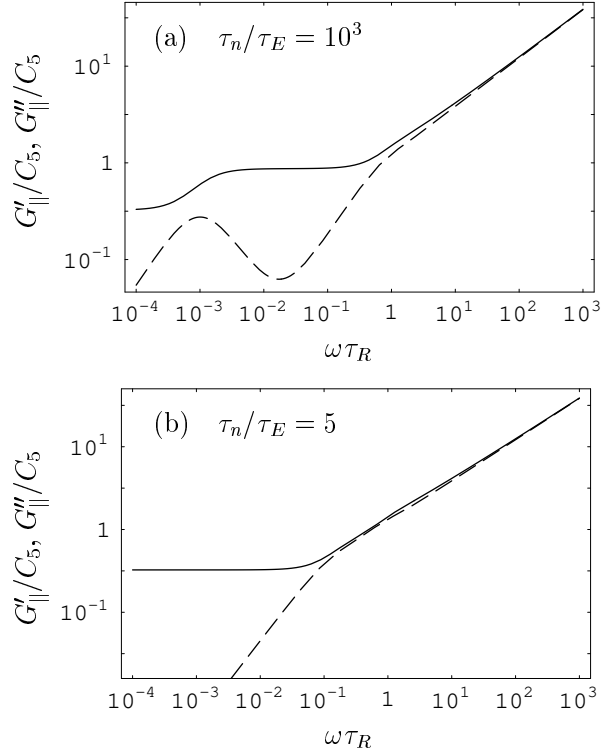}
\end{center}
\caption[]{\label{r_5_1000}Log-log plot of the reduced storage
and loss moduli $G_{||}^\prime/C_5$ (solid lines) and
$G_{||}^{\prime \prime}/C_5$ (dashed lines) versus the
reduced frequency $\omega \tau_E$ as given respectively by
the real and imaginary parts of Eq.~(\ref{complexModulus})
(a) for $\tau_n/\tau_E=10^3$ and (b) for $\tau_n/\tau_E=5$.
For the purpose of illustration we have set, by and large
arbitrarily, $\nu_5 / (C_5 \tau_E) = 1.8$, $D_1/C_5 =
0.0135$, $D_2/D_1 = 10$ and $\lambda = -1$.}
\end{figure}

The phenomenological equations for $\nv$ and $\uv$, which
can be derived using standard Poisson-bracket
approaches \cite{forster&Co_71_Forster1983},
are \cite{stenull_lubensky_2004}
\begin{subequations}
\label{genStruct}
\begin{eqnarray}
\label{EOMa} \dot{n}_i &=& \lambda_{ijk} \, \partial_j
\dot{u}_k - \Gamma \, \frac{\delta \mathcal{H}}{\delta
n_i}\, ,
\\
\label{EOMc} \rho \ddot{u}_i &=& \lambda_{kji} \, \partial_j
\frac{\delta \mathcal{H}}{\delta n_k}  - \frac{\delta
\mathcal{H}}{\delta u_i} + \nu_{ijkl}\, \partial_j
\partial_l \dot{u}_k \, ,
\end{eqnarray}
\end{subequations}
where $\nu_{ijkl}$ is an unrenormalized uniaxial viscosity
tensor with five independent components, $\Gamma$ is a
dissipative coefficient with dimensions of an inverse
viscosity, and
\begin{eqnarray}
\lambda_{ijk} = \frac{\lambda}{2} \, \left( \delta_{ij}^T
\, n_k  + \delta_{ik}^T \,n_k \right)+ \frac{1}{2} \,
\left( \delta_{ij}^T \, n_k  - \delta_{ik}^T \,n_k \right)
,
\end{eqnarray}
where $\delta_{ij}^T = \delta_{ij} - n_i n_j$. The complex
modulus $G^\ast_\parallel (\omega) = C_5^R (\omega)$
predicted by these equations is
\begin{equation}
\label{dynRen}
G^\ast_\parallel (\omega)= C_5 - i \omega \left( \nu_5 +
\frac{\lambda^2}{2\, \Gamma} \right) - \frac{D_2^2}{2\,
D_1} \, \, \frac{(1 - i \omega \tau_2)^2}{1-i \omega
\tau_1},
\end{equation}
where $\tau_1 = 1/(\Gamma D_1)$ and $\tau_2 =
-\lambda/(\Gamma D_2)$. This result is equivalent to the
one obtained by Terentjev, Warner and
coworkers~\cite{WarnerTer2003,terentjev&Co_NEhydrodyn}.
Equation~(\ref{dynRen}) is a valid expression for the
complex modulus provided $\tau_1 \equiv \tau_n$ is much
longer than any other nonhydrodynamic relaxation time,
including elastomer times, of the system. As can be seen in
Fig.~\ref{r_5_1000}(a), it predicts a plateau in
$G'(\omega)$ for $\omega > \tau_n^{-1}$.

In standard polymeric systems, the Rouse time, $\tau_R$, is
of order $10^{-5}$s, and light scattering measurements on
nematic elastomers yield $\tau_n \sim
10^{-2}$~\cite{schoenstein&Co_2001}. It could, therefore,
be argued that $\tau_E \sim \tau_R$ and that that $\tau_n
\gg \tau_E$. However, liquid crystalline polymers are more
viscous than standard polymers, and it is not unreasonable
for $\tau_E$ to be considerably less than $10^{-5}$s.  If
$\tau_n$ and $\tau_E$ are of the same magnitude, or if
$\tau_n < \tau_E$, then the viscosities and decay times
$\tau_1$ and $\tau_2$ in Eq.~(\ref{dynRen}) must be treated
as having a frequency dependence when $\omega \tau_E$ is of order 1 or greater.

Since high-frequency dynamics of the nematic phase are
expected to be similar to that of the nearly isotropic
paranematic phase, $\nu_5$ should have the same behavior as the
viscosity in the latter phase:
\begin{align}
\nu_5 (\omega)= \nu_5 f_R ( - i \omega \tau_E).
\label{eq:nu5}
\end{align}
In addition, the elastomer modes will affect the relaxation
of director modes. Indeed, the light scattering experiments
of Ref.~\cite{schoenstein&Co_2001} indicate a very broad
distribution of director relaxation times. There is no
reason why many closely spaces modes should not lead to a
behavior of $\Gamma^{-1}$ with a Rouse-like form,
\begin{align}
\Gamma (\omega)^{-1}= \Gamma^{-1} f_R ( - i \omega \tau_E).
\label{eq:Gamma}
\end{align}
With this assumption, a reasonable phenomenological form
for the complex modulus $G_{||}^*(\omega)$ that
incorporates the effects of both Rouse and director modes
is
\begin{align}
\label{complexModulus}
& G_{||}^*(\omega) = C_5 + \left(\frac{\nu_5}{\tau_E } +
\frac{\lambda^2}{2}
\frac{\tau_n}{\tau_E}\, D_1\right) h_R ( - i\omega \tau_E ) \nonumber \\
&  - \frac{D_1}{2}\, \left(\frac{D_2}{D_1}\right)^2
\frac{[1- \lambda(D_1/D_2)(\tau_n/\tau_E) \, h_R ( - i
\omega \tau_E)]^2}{1 + (\tau_n/\tau_E)\, h_R(-i \omega
\tau_E )} \, .
\end{align}
$G_{\perp}^* ( \omega )$, on the other hand, has the same
form as for an isotropic system, cf.\ Eq.~(\ref{eq:G*2}).
The storage and loss moduli $G_{||}^\prime (\omega)$ and
$G_{||}^{\prime \prime} (\omega)$ implied in
Eq.~(\ref{complexModulus}) are plotted in
Fig.~\ref{r_5_1000}. If $\tau_n \leq \tau_E$ the frequency
dependence of the moduli is dominated by the elastomer
modes with $G_{||}^\prime (\omega) = C_5^R$ and
$G_{||}^{\prime \prime} (\omega) = \nu_5^R \omega $ with
$\nu_5^R = \nu_5 + [\lambda^2/(2 \Gamma)]
(1-\tau_1/\tau_2)^2$ for $\omega \tau_E \ll 1$ and
$G_{||}^\prime (\omega) \sim G_{||}^{\prime \prime}
(\omega) \sim \sqrt{\omega}$ for $\omega \tau_E \gg 1$.  If
$\tau_n \gg \tau_E$, however, the structure is much richer.
In addition to the behavior for $\tau_n \leq \tau_E$, the
storage and loss moduli develop respectively a noticeable
plateau and dip at intermediate frequencies.
%%%%%%%%%%%%%
%%%%%%%%%%%%%

Two experimental groups, the Cambridge
group~\cite{clarke&Co_2001,hotta_terentjev_2003,terentjev&Co_2003}
and the Martinoty group~\cite{stein&Co_2001,martinoty&Co_2004}, have reported
extensive measurements of $G_{||}^\ast$ and
$G_{\perp}^\ast$ on a variety of samples. The paper of
Martinoty group shows curves of $G_{\parallel}^{\prime} (
\omega)$ and $G_{\parallel}^{\prime \prime} ( \omega)$ that
are very similar to those in Fig.~\ref{r_5_1000}(b) for
$\tau_n/\tau_E = 5$. Remarkably, the experimental curves
are also compatible Fig.~\ref{r_5_1000}(a) where
$\tau_n/\tau_E \ll 1$ if one ignores the low-frequency part
of the theoretical curves.

In their simplest form in which $\Gamma$ is independent of
frequency, the NED equations treat the non-hydrodynamic
director modes but no other non-hydrodynamic modes.
Ignoring other non-hydrodynamic modes is only justified if
there is a clear separation of time scales with $\tau_1 \gg
\tau_E$. Martinoty {\em et al}.\ argue convincingly that
their data is incompatible with this assumption.  Indeed
the Cambridge group does not disagree with this observation
at least in some materials since in
Ref.~\cite{terentjev&Co_2003}, they indicate explicitly
that there is insufficient separation of time (actually
frequency) scales for the intermediate plateau shown in Fig.~\ref{r_5_1000}(a) to be observed.  Our view is that the experimental evidence is that $\tau_n$ is of
order $\tau_E$ in current experimental systems, though
current data does not rule out the possibility of $\tau_n$
being as much as $5$ times greater than $\tau_E$ as a
comparison of Fig.~\ref{r_5_1000}(b) and the Martinoty data
shows. Though it seems unlikely that $\tau_n \gg \tau_E$,
one cannot even rule out this possibility entirely on the
basis of the current experimental data, because the
frequencies in the experiments might not be low enough to
probe $\tau_n^{-1}$ if $\tau_n$ is greater than several
hundred seconds. As a comparison of the experimental curves and Fig.~\ref{r_5_1000}(a)
shows, there is a danger of overlooking the intermediate
plateau in $G_{||}^{\prime}$ and the dip in $G_{||}^{\prime
\prime}$ if frequencies less than $\tau_n^{-1}$ are not
probed.

The measurements by the Cambridge group and the Martinoty
group include 2 samples each prepared in the same way by
both groups. For these samples, the curves for
$G_{\perp}^{\prime}$ as a function of temperature $T$ at
different $\omega$ are very similar.  At each frequency,
$G_{\perp}^{\prime}$ shows a general upward trend
with decreasing $T$ as the glass transition is approached.
At the lowest frequency, there is a measurable dip in the
vicinity of the $IN$ transition, though that reported by
the Martinoty group is smaller than that reported by the
Cambridge group. The magnitude of this dip decreases with
increasing frequency. Both groups also report a decrease in
$G_{||}^{\prime}/G_{\perp}^{\prime}$ at the $IN$
transition.

To an impartial reader, the data obtained by the two groups
are qualitatively if not quantitatively similar. The two
groups, however, provide different interpretations of their
data. The Cambridge group interprets the dip in
$G_{||}^{\prime}/G_{\perp}^{\prime}$ as a manifestation of
dynamic softening associated with the transition from
uniaxial paranematic phase to the semi-soft nematic phase
whereas the Martinoty group says that the dip is simply
explained in terms of the de Gennes model, in which the
reduction of $G_{||}$ arises from the $D_2^2/D_1$ term,
without recourse to the concept of semi-soft elasticity.

The de Gennes theory for nematic elastomers defined by the
total elastic energy $\mathcal{H}$ provides a correct
description of any uniaxial elastomer regardless of how it
is formed.  A soft uniaxial elastomer formed via
spontaneous symmetry breaking from an isotropic elastomer
is merely one with $C_5^R = C_5 - D_2^2/(2D_1) = 0$.  Since
the high-temperature phase of single-domain nematic
elastomers are at least weakly uniaxial, it is true that
all single-domain phases can be described by the de Gennes
model without invoking the concepts of soft or semi-soft
elasticity.  This approach, however, misses the fact that
the vanishing of $C_5^R$ and the resultant soft elasticity
in nematic phases formed spontaneously from isotropic
elastomers is a consequence of symmetry and not of
arbitrary juggling of parameters.  A transition from a
weakly uniaxial paranematic phase to a more strongly
uniaxial nematic phase will be accompanied by a decrease in
$C_5^R$ that increases as the anisotropy of the paranematic
phase decreases.  In the resultant semi-soft nematic phase,
$C_5^R$ vanishes with the anisotropy of the paranematic
phase.  Our preference is to use the concepts of soft and
semi-soft elasticity to describe the reduction of $C_5^R$
that arise from the paranematic-to-nematic transition. We
agree, however, that these concepts become less meaningful
in samples that are really above the mechanical critical
point~\cite{de_Gennes-2} at which this transition becomes
continuous.  Being above this critical point is, however,
incompatible with the large plateau in the semi-soft
non-linear stress strain curve~\cite{Kupfer}. The Martinoty
data does seem to be consistent with being above the
mechanical critical point.  If it is, them we would like to
see experiments done on materials that are softer in the
nematic phase if they really do exist.

The dynamics of nematic elastomers are clearly very
complicated.  They have all of the complexity of rubbers
and gels, which have localized gel and other modes that we
have not even discussed and eventual transitions to glassy
phases, in addition to that of nematics. In spite of these
complexities, the experimental results for the complex
storage modulus of a number of samples of nematic
elastomers measured by two groups are remarkably similar.
Neither group finds a truly soft nematic phase with a shear
modulus an order of magnitude or more smaller than it is in
the high-temperature nearly isotropic phase, but their data
can be described in the context of semi-soft elasticity.

The above discussion summarizes what current theories can
say about the equilibrium rheological properties of
homogeneous nematic elastomers.  The only rigorous
statement is that semi-soft nematic elastomers are uniaxial
solids with frequency-independent shear moduli and
viscosities in the low-frequency limit.  At higher
frequencies, contributions from non-hydrodynamic modes,
including director relaxation, contribute to the complex
moduli.  Though experiments seem to be converging to
reproducible results, it is clear that there is much more
to do before we have a complete understanding of the
dynamics of nematic elastomers.

We gratefully acknowledge support by the National Science
Foundation under grant DMR 0404670.


\begin{thebibliography}{}
\bibitem{footnote1}
We define temporal Fourier transforms via
$A(t) = (2\pi)^{-1} \int d \omega \, A (\omega) \exp (-i \omega t)$ for any function $A(t)$.
\end{thebibliography}
\end{document}